\documentclass[aps,prd,floats,superscriptaddress, showpacs]{revtex4}
\usepackage{graphicx}
\newcommand{\newc}{\newcommand}
\newc{\beq}{\begin{equation}}
\newc{\eeq}{\end{equation}}
\newc{\beqa}{\begin{eqnarray}}
\newc{\eeqa}{\end{eqnarray}}
\newc{\bray}{\begin{array}}
\newc{\eray}{\end{array}}
\newc{\IM}{\mbox{\sl{Im}}}
\newc{\RE}{\mbox{\sl{Re}}}
\newc{\bit}{\begin{itemize}}
\newc{\eit}{\end{itemize}}
\newc{\CC}{\mathbf{c}}
\newc{\nonr}{\nonumber}
\newc{\hs}{\hskip 3mm}
\newc{\ra}{\rightarrow}
\newc{\TR}{\mbox{\sl{Tr}}}
\newc{\tri}{\triangle}
\newc{\MS}{M^*}
\newc{\VEVB}{\tilde{v}_b}
\newc{\bi}{\begin{itemize}}
\newc{\ei}{\end{itemize}}
\newc{\trip}{$\mathbf{3}$ }
\newc{\sext}{$\bar{\mathbf{6}}$ }
\newc{\tripp}{$\mathbf{3^{\prime}}$ }
\newc{\xw}{\sin^2\theta_{\rm W}}
\begin{document}

\pagestyle{plain}

\title{Lepton Flavor Violation in Extra Dimension Models}

\author{We-Fu Chang}
\email{wfchang@phys.sinica.edu.tw}
\affiliation{Institute of Physics, Academia Sinica, Taipei 115, Taiwan}
\affiliation{TRIUMF Theory Group,
4004 Wesbrook Mall, Vancouver, B.C. V6T 2A3, Canada}

\author{John N. Ng}
\email{misery@triumf.ca}
\affiliation{TRIUMF Theory Group,
4004 Wesbrook Mall, Vancouver, B.C. V6T 2A3, Canada}
\date{\today}

\begin{abstract}
Models involving large extra spatial dimension(s) have interesting predictions
on lepton flavor violating processes. We consider some 5D models which
are related to neutrino mass generation or address the fermion masses hierarchy problem.
 We study the signatures in low energy experiments that can  discriminate the  different models.
The focus is on   muon-electron conversion in nuclei,
$\mu\ra e \gamma$  and $\mu\ra 3e$ processes and their $\tau$ counterparts.
Their links with the active  neutrino mass matrix are investigated.
We show that in the  models we discussed  the branching ratio of
$\mu\ra e \gamma$ like rare process is much smaller than the
ones of $\mu\ra 3e$ like processes. This  is in sharp contrast  to most
of the traditional wisdom based on four dimensional gauge models.
Moreover, some rare tau decays are more promising than the rare muon
decays.

\end{abstract}
\maketitle

\section{Introduction}
\label{sec:intro} In the Standard Model(SM) with fifteen fermions per family neutrinos are
strictly massless and  the  charged leptons' weak eigenstates
can be chosen to be their mass eigenstates. Thus, each  generation
has a separately conserved lepton number.  If one neglects the
tiny effects from nonperturbative processes, there is no lepton
flavor violating(LFV) interaction in SM. However, recent
neutrino experiments show strong evidence that neutrinos have
none zero masses and the three active neutrinos  mix\cite{SuperK,SNO,KamL,WMAP}. Most
physicists take this to be a harbinger  of new physics beyond the SM. Moreover, finite
neutrino masses alone would  imply  the existence  of LFV in
charged lepton sector analogous to the quarks.
If so  we expect the Glashow-Iliopoulos-Maiani(GIM) mechanism to be operative in the lepton sector
 then  the rate of induced LFV processes  will be proportional to the neutrino
mass square difference, which is of the order of $<10^{-3}
(eV)^2$. Hence, they will be hopelessly  small for experimental verification. Therefore, additional
ingredients are essential for a detectable LFV signature. It is very common in model building to have
the new physics that generate neutrino masses also given rise
to LFV reactions. This link appears to be natural although there is no guarantee that this is case
in nature. With this cautionary note we will focus attention to new physics that links the two
phenomena.

Among the numerous beyond the SM models, LFV signatures are most intensely
studied in supersymmetric (SUSY) ones. The connection with neutrino masses is established through
the seesaw mechanism which is the orthodox way of getting a small mass for the active neutrinos.
Since the latter has a natural setting in grand unified theories (GUT) the end result
are  rather bedecked  supersymmetric seesaw models; see e.g. \cite{susylfv}.
Although the details are different the generic source of LFV lies in the mixing of
various sfermions. The right-handed Majorana neutrinos play a secondary role in this class
of models. In general
it is natural to expect $B(\mu\ra e \gamma)\gg B(\mu\ra 3e)$ in SUSY models.

For non-supersymmetric models neutrino mass generation via the seesaw mechanism would
 required the right-handed neutrinos to be of the GUT scale. In this simplest version
all LFV are undetectable. Attempts are now made to lower some right-handed neutrinos mandated by
the seesaw mechanism to the TeV scale so that the seesaw mechanism itself can be tested
experimentally. If so then one can optimistically anticipate LFV signatures in the next round of
experiments \cite{CN04}. Independent of the details of the models  one again expects $B(\mu\ra e \gamma)\gg B(\mu\ra 3e)$
to hold true.

Recently a new avenue has open up in the construction of models beyond the SM that
exploits the possible existence of extra spatial dimensions. These theories are
particularly interesting phenomenologically in the brane world context.
It is fascinating that many
long standing problems in the usual four dimensional (4D) field theories can be overcome or
take on new perspectives in these higher dimensional constructs. For example the hierarchy
problem is solved by invoking large extra dimensions.
In this note, we would like to draw the readers' attention to the
models which involve one or more flat extra spatial dimensions. Furthermore, we focus on those that
address the neutrino mass problem. In some cases, we predict a
reversed pattern of  $B(\mu\ra 3e)\gg B(\mu\ra e \gamma)$ compare to SUSY models.  On the
experimental side, it shall be interesting to see this.

The current experimental limits on muon LFV  have already put very
stringent constraints on model building. On the other hand, the
limits from  tau LFV are  rather loose.  We shall constraint the
extra dimension  models  by the  muon rare processes data
and place upper limits on the rare decays of the $\tau$. To avoid any
 hadronic uncertainties we shall focus on purely leptonic processes.
We shall also  discuss the possible ways to discriminate
different models and their connections to neutrino masses.

In this brief review, we give few examples of extra dimension
models which give potentially testable LFV signatures.
These LFV processes are all directly or indirectly  related to the generation of
neutrino masses. We compare the LFV processes in a
five dimension(5D) $SU(3)_W$\cite{SU3:triumf}
and $SU(5)$\cite{SU5:triumf} GUT models where neutrino Majorana
masses are generated radiatively without a right-handed neutrino which is viable
but less discussed alternative to the seesaw mechanism. A brief review
of this construction is  given in \cite{JN}.
Also, we discuss the  LFV processes in split fermion or multi-brane
scenario.

The  following is our plan for the paper.
In sec.II, we will first review the general operator analysis for
the lepton flavor violating processes. This will also set the notations
for the rest of the discussions.
Sec. III examines  LFV in a 5D $SU(3)_w$ model. New results of the one-loop calculations
are given here. For details of the model and neutrino mass generation we refer to
\cite{SU3:triumf}.
In Sec. IV, the LFV processes induced by 5D $SU(5)$ model will be
discussed. The discussion here has not been presented before.
The alternative way of studying the flavor problem using the split fermion model is examined in Sec.V.
Calculations of the LFV processes in this scenario involves many new unknown parameters. The models
lack  predictive power even semi-quantitatively. However, very general generic trends for LFV can be
discerned even in this early stages of development.
Our conclusions are given in Sec. VI.
The necessary 5D gauge fixing details, which is crucial for loop
calculations, are  presented in an appendix.

\section{General operator analysis}
First of all, we collect the necessary general formulas for the study of
LFV processes. The most important ones are the effective interactions of $L-l-\gamma$
and $L-l-Z$
where we use the notation $L$ to denote the heavier charged lepton which usually is either
$\mu$ or $\tau$
and $l$ is the lighter daughter lepton which can be $\mu$  or $e$.

In LFV studies, the most important contribution comes from the effective $L-l-\gamma$
vertex. The similar vertex where a virtual $Z$ replaces the $\gamma$ is subdominant in the class
of models we are considering. For definiteness we will take $L= \mu$ and $l= e$.
The most general $\mu-e-\gamma$ interaction amplitude allowed by
Lorentz and gauge invariance can be written as: \beq {\cal M}= -e
A^*_\mu(q) \overline{u_e}(p_\mu-q)\left\{ \left[
f_{E0}(q^2)+f_{M0}(q^2) \gamma^5\right]\gamma_\nu
\left(g^{\mu\nu}-{q^\mu q^\nu \over q^2} \right) + \left[
f_{M1}(q^2)+f_{E1}(q^2) \gamma^5\right]{i\sigma^{\mu\nu}q_\nu
\over m_\mu} \right\}u_\mu(p_\mu) \eeq with the convention  $e=|e|>0$
used  through out this paper and $q^{\mu}$ is the photon 4-momentum. For real photon emission, only
$f_{E1}$ and $f_{M1}$ contribute. But if a off-shell photon is
involved, then  all 4 form factors contribute. After proper
renormalization, the amplitude is finite  as $q^2\ra 0$, so
we must have
$f_{E0}(0)=f_{M0}(0)=0$. It is customary to  factor out $q^2$ and rewrite
the electric and magnetic form factors as
\beq \label{eq:mu_e_form} f_{E0}(q^2)={q^2\over
m_\mu^2}\tilde{f}_{E0}(q^2)\,,\, f_{M0}(q^2)={q^2\over
m_\mu^2}\tilde{f}_{M0}(q^2)
\eeq
 and now $\tilde{f}_{E0}(q^2)$ and
$\tilde{f}_{M0}(q^2)$ are finite at $q^2\ra 0$.

\subsection{ $L\ra l_1 l_2 \bar{l}_3$ and $ L\ra l \gamma$}
Using similar notations of \cite{Okada}, the most general
effective lagrangian for $\mu\ra 3e$ and $\mu\ra e \gamma$ can be
expressed as: \beqa -{\sqrt{2}{\cal L}\over 4 G_F  }&=& m_\mu A_R
\overline{e_R} \sigma^{\mu\nu}\mu_L F_{\mu\nu} + m_\mu A_L
\overline{e_L} \sigma^{\mu\nu}\mu_R F_{\mu\nu} + g_1
(\overline{e_R}\mu_L)(\overline{e_R}e_L) + g_2
(\overline{e_L}\mu_R)(\overline{e_L}e_R)\nonr\\ &+& g_3
(\overline{e_R}\gamma^\mu \mu_R)(\overline{e_R}\gamma_\mu e_R) +
g_4 (\overline{e_L}\gamma^\mu \mu_L)(\overline{e_L}\gamma_\mu
e_L)\nonr\\ &+& g_5 (\overline{e_R}\gamma^\mu
\mu_R)(\overline{e_L}\gamma_\mu e_L) + g_6
(\overline{e_L}\gamma^\mu \mu_L)(\overline{e_R}\gamma_\mu e_R) +
h.c. \eeqa where \beq A_R=  -{\sqrt{2} e \over 8 G_F
m_\mu^2}\left[f_{E1}(0)+f_{M1}(0)\right]\,,\, A_L=  -{\sqrt{2} e
\over 8 G_F m_\mu^2}\left[f_{M1}(0)-f_{E1}(0)\right]. \eeq Also
note the anapole form factors $f_{E0}$ and $f_{M0}$ have  vector like
effective contributions to $g_{3-6}$:
 \beqa
  \delta g_3 = \delta g_5=  {\sqrt{2} e^2 \over
4G_F m_\mu^2}\left[\tilde{f}_{E0}(0) - \tilde{f}_{M0}(0)\right]\\
\delta g_4 = \delta g_6= {\sqrt{2} e^2 \over 4G_F
m_\mu^2}\left[\tilde{f}_{E0}(0) + \tilde{f}_{M0}(0)\right]
\eeqa
which shall be included in  the $g_{3,4,5,6}$. The above effective
lagrangian leads to
\beqa
B(\mu\ra e\gamma)&=& 384\pi^2
(|A_L|^2+|A_R|^2)\\ \label{eq:taudecay} B(\mu\ra
3e)&=&{|g_1|^2+|g_2|^2\over 8} +
2(|g_3|^2+|g_4|^2)+|g_5|^2+|g_6|^2\nonr\\ &&+ 8e Re\left[A_R(
2g_4^*+g_6^*)+A_L (2g_3^*+g_5^*)\right]\nonr\\ &&+ 64e^2
\left\{\ln\frac{m_\mu}{m_e}-\frac{11}{8}\right\}(|A_R|^2+|A_L|^2)
\eeqa
if electron mass is ignored.

To carry out the calculation, it's convenient to define two
dimensionless variables
$x_1=2E_1/m_\mu$ and $x_2=2E_2/m_\mu$.
However, it is important to keep $m_e^2$ terms in the intermediate steps
in order to properly extract the finite term in the last line
of Eq.(\ref{eq:taudecay}). Our result agrees with \cite{Petcov,Okada}.

The expressions of Eq.(\ref{eq:taudecay}), except the last line of dipole operators,
can also apply to $\tau\ra l+\gamma$ and
$\tau\ra l_1 l_1 \overline{l_3}$ processes .
For $\tau\ra e e\bar{\mu}, \mu \mu\bar{e}$ processes, the part from dipole operators
have double loop suppression from two flavor violation vertices and resulting in an
insignificant branching ratios thus  can be safely ignored.
For $\tau$ decay, the branching ratios given above are  normalized to $B(\tau\ra e
\bar{\nu_e}\nu_\tau)$. This holds for subsequent discussions of $\tau$ decays.

To complete the story, we also give the expression for processes
with no identical particles in the final state,
namely, $\tau\ra\mu \bar{e} e, e \mu\bar{\mu}$ or $l_1 \neq l_2=l_3$.
The above expression for the branching ratio is now modified to:
\beqa
\label{eq:taudecay2}
B(\tau\ra l_1 l_2 \overline{l_3})&=&{|g_1|^2+|g_2|^2\over 4} +
(|g_3|^2+|g_4|^2+ |g_5|^2+|g_6|^2)\nonr\\
&&+ 8e Re\left[A_R( g_4^*+g_6^*)+A_L (g_3^*+g_5^*)\right]\nonr\\
&&+ 64e^2
\left\{\ln\frac{m_\tau}{m_2}-\frac{3}{2}\right\}(|A_R|^2+|A_L|^2)
\eeqa
with trivial extension of $g_i$s.
In arriving the last line of Eq.(\ref{eq:taudecay2}), we have
ignored the masses difference between $m_e$ and $m_\mu$ in phase space integration
but keep the crucial mass singularity associated with the virtual photon.
Not surprisingly, the approximation agrees very  well with the actual numerical
integrations.

If the photonic dipole operator is the only dominate LFV source,
we have the following  model independent prediction for
\beqa
B(\tau\ra e\gamma) &=&  B(\tau \ra \mu \gamma)\\
{B(\tau \ra \mu e\bar{e})\over B(\tau\ra e\gamma) } &=&
\frac{2\alpha}{3\pi}\left\{ \ln\frac{m_\tau}{m_e}-\frac32
\right\}\\
{B(\tau \ra e \mu\bar{\mu})\over B(\tau\ra e\gamma) } &=&
\frac{2\alpha}{3\pi}\left\{ \ln\frac{m_\tau}{m_\mu}-\frac32
\right\}
\eeqa
to the accuracy of $m_{\mu}^2/m_{\tau}^2$.
To our knowledge, the last two relations have not been presented before.

\subsection{$\mu-e$ conversion in nuclei}
We can write  the effective LFV Lagrangian  for $\mu-e$ conversion as:
\beqa
{{\cal L}_{\rm eff} \over \sqrt{2}G_F} &=&
 \bar{e}\left(s -p\gamma^5\right) \mu \sum_q  \bar{q}\left(s_q-p_q\gamma^5\right)q
+ \bar{e} \gamma^\alpha\left(v-a\gamma^5\right)\mu \sum_q \bar{q}
\gamma_\alpha\left(v_q-a_q\gamma^5\right)q\nonr\\
&+& \frac12 \bar{e}\left(t_s + t_p \gamma^5\right)\sigma^{\alpha\beta}\mu
\sum_q \bar{q} \sigma_{\alpha\beta} q + H.c.
\label{eq:LmueTi}
\eeqa
with self explanatory notations.
Here, flavor changing terms in the quark sector are  not included since they are not
expected to be important here.
The effective couplings are normalized to
$(\sqrt{2}G_F)^{1/2}$. For example,  the SM $Z$ boson has a vector
coupling to quarks given by
\[
v^q=T_3-2Q\sin^2\theta\,.
\]

To calculate the conversion rate, we need to promote the
interaction from quark level to the  nucleon level by computing the
matrix elements $\left\langle N|\bar{q}\Gamma q | N\right\rangle = G_\Gamma^{q,N}
\bar{N}\Gamma N$ with $N$ denotes a nucleon and
$\Gamma=\{1,\gamma^5, \gamma_\alpha, \gamma_\alpha\gamma^5,
\sigma_{\alpha\beta}\}$.
Since the coherent process is the important one  only vector and scalar operators matter:
\beq
\left\langle p|\bar{q}\gamma_\alpha q | p\right\rangle = G_V^{q,p}
\bar{p}\gamma_\alpha p\,,\,
\left\langle n|\bar{q}\gamma_\alpha q | n\right\rangle = G_V^{q,n}
\bar{n}\gamma_\alpha n
\eeq
and
\beq
\left\langle p|\bar{q} q | p\right\rangle = G_S^{q,p}\bar{p} p\,,\,
\left\langle n|\bar{q} q | n\right\rangle = G_S^{q,n}\bar{n} n\,.
\eeq
By conserving of vector current, in the $q^2\sim 0$ limit, one can
determine that $G_V^{u,p}=G_V^{d,n}=2$ and $G_V^{u,n}=G_V^{d,p}=1$.
However, one has to rely on the nucleon model to evaluate the scalar
operator. For qualitative estimation, we will use the result $G_S \sim G_V$
from full non-relativistic quark model but the reader should keep in mind
that the uncertainty of nucleon model could be as large as few tens percent\cite{Huitu}.
Following the approximations used  in \cite{weinberg}, the conversion
rate, normalized to the normal muon capture rate $\Gamma_{\rm capt}$,
can be expressed as\cite{weinberg,Okada, Kitano:2002mt}:
\beq
\label{eq:mu-e-barnchrate}
B_{\rm conv}=
{p_e E_e G_F^2 F_p^2 m_\mu^3 \alpha^3 Z_{eff}^4 \over 2 \pi^2 Z \Gamma_{\rm capt}}
\left\{ |4 e A_L Z  +(s-p)S_N +(v-a)Q_N|^2
+ |4 e A_R Z  +(s+p)S_N +(v+a)Q_N|^2 \right\}
\eeq
by assuming that the proton and neutron density are
equal and the muon wave function does not change very much in the
nucleus, and $F_p$ is a form factor whose definition can be found in \cite{weinberg}
and $p_e (E_e)$ is the electron momentum (energy), $E_e\sim p_e\sim m_\mu$.
For $^{48}_{22}Ti(^{27}_{13}Al)$, $F_p \sim 0.55(0.66)$, $Z_{eff}\sim
17.61(11.62)$, and $\Gamma_{capture}\sim 2.6(0.71)\times 10^{6}
s^{-1}$\cite{Chiang:1993xz}.

Where the coherent vector and scalar coupling strength of nuclei
$N$ are defined as
\beqa
S_N &\equiv& s^u(2Z+N)+s^d(2N+Z)\,,\\
Q_N &\equiv& v^u(2Z+N)+v^d(2N+Z).
\eeqa
If there are more than one gauge or scalar bosons mediate this
process, the above expression can be trivially extended with modified
couplings:
\beqa
(s\pm p)S_N \Rightarrow \sum_i (s^i \pm p^i) S^i_N
\frac{M_Z^2}{M_{Hi}^2}\,,\\
(v\pm a)Q_N \Rightarrow \sum_i (v^i \pm a^i) Q^i_N
\frac{M_Z^2}{M_{Zi}^2}\,.
\eeqa
Note that the form factors $\tilde{f}_{E0}$ and $\tilde{f}_{M0}$ in
Eq.(\ref{eq:mu_e_form}) have extra  contribution to the vector couplings:
\beqa
\delta v= -{2 e M_W \over g m_\mu}\tilde{f}_{E0}\,,\
\delta a= -{2 e M_W \over g m_\mu}\tilde{f}_{M0}\,,\
\delta v_q= {2 e M_W \over g m_\mu}Q_q\,,
\eeqa
and if Eq.(\ref{eq:mu_e_form}) is the only LFV source, then Eq.(\ref{eq:mu-e-barnchrate}) reduces
to the well-known formula given in \cite{weinberg}
\beq
B^{\gamma}_{\rm conv}=
{8 m_\mu F_p^2 \alpha^5 Z_{eff}^4 Z \over \Gamma_{\rm capt}}
\left\{|f_{M1}+f_{E0}|^2 + |f_{M0}+f_{E1}|^2\right\}.
\eeq
Also a model-independent relation between the $\mu-e$
conversion and the $\mu\ra e\gamma$
\beq
\label{eq:Weinberg}
B^{\gamma}_{\rm conv} = { m_\mu^5 G_F^2 F_p^2 \alpha^4 Z_{eff}^4 Z \over 12\pi^3\Gamma_{\rm capt} }
\left({ |f_{M1}+f_{E0}|^2 + |f_{M0}+f_{E1}|^2 \over |f_{M1}|^2 + |f_{E1}|^2
}\right)
B(\mu\ra e+\gamma)\,.
\eeq
The above brief review  is sufficient for the  phenomenological
analysis we do.  Next, we will head for the extra-dimensional models and discuss their LFV signatures.

\section{5D $SU(3)_W$ unification model}

It has been known for a long time that the SM lepton left-handed doublet
and the right-handed singlet charged lepton in  each family
can  beautifully form an  $SU(3)_W$ fundamental
representation, i.e. $ L=\left( e,  \nu, e^c \right)^T_{\mathrm L}$ \cite{Wein75}.
This is implemented in an electroweak only unification in which $SU(2)\times U(1)$
is unified  to $SU(3)_W$.
One of the  attractive points  of this  unification model
is the tree level prediction of $\sin^2\theta_W=1/4$. Renormalization group
considerations point to a
relatively  low scale of unification at $\sim$ few $TeV$.
We shall use $\{U^{\pm2}, V^\pm\}$ to denote
the $SU(3)_W/(SU(2)\times U(1))$ gauge bosons
which have SM quantum number $(2,\pm 3/2)$.
In 4D, the $SU(3)_W$ GUT has a fundamental
difficulty of embedding quarks into $SU(3)_W$
representations.
This problem can be circumvented  by promoting the model into
five dimensional space time \cite{5DSU3} and \cite{SU3:triumf}.
We give a brief summary of the model construction here.

The  extra spatial dimension, with coordinate denoted by $y$,
is compactified into an $S_1/(Z_2\times Z'_2)$ orbifold.
Where the circle  $S_1$ of  radius $R$, or $y=[-\pi R, \pi R]$,
is orbifolded by a $Z_2$ which identifies points $y$ and $-y$.
The resulting space is further divided by a second $Z_2^{\prime}$ acting
on $y'= y - \pi R/2$ to give the final geometry.

We now have tow parity transformations
 $P: y\leftrightarrow -y$ and $P': y'\leftrightarrow -y'$ under which the
bulk fields can be assigned either of the eigenvalues + or -. This freedom is
used to break the
 bulk $SU(3)_W$ symmetry to  $SU(2)\times U(1)$. Explicitly, one
assigns   the following properties to bulk gauge fields
\beqa
{\cal A}_\mu(y) = {\rm P} {\cal A}_\mu(-y) {\rm P}^{-1}\,,&&
{\cal A}_\mu(y') = {\rm P}' {\cal A}_{\mu}(-y') {\rm P}'^{-1}\nonr\\
{\cal A}_5 (y) = -{\rm P} {\cal A}_5(-y){\rm P}^{-1}\,,&&
{\cal A}_5(y') = -{\rm P}' {\cal A}_5(-y'){\rm P}'^{-1}
\eeqa
where the  matrices ${\rm P}=diag\{+++\}$ and
${\rm P}'=diag\{++-\}$.
Now  the $(Z_2,Z'_2)$ parities of the SM gauge bosons
and the $U,V$ gauge bosons  are $(++)$ and $(+-)$ respectively.
It is  easy to work out the Fourier eigenmodes propagating in the bulk and see that
only fields with $(++)$ parity have zero modes.
In other words, only SM gauge bosons have zero modes. Both the $U,V$ gauge bosons
and all the $y-$components are heavy KK excitation. Note the
second $Z_2'$ is necessary to avoid the presence of zero modes for both
SM gauge boson and the exotic $U^{\pm2}, V^\pm$ boson at the same
time.

The $SU(3)_W$ symmetry is explicitly broken to $SU(2)_L\times
U(1)$ at  the $y=\pi R/2$ fixed point, where
the  4D quarks field are forced to live on it.
The extra degree of freedom in extra dimensional theories
is the key to incorporate SM quarks into the $SU(3)_W$ symmetry.
On the other hand, the lepton fields can be placed anywhere in the bulk or on either
two fixed points.
We choose to put the 4D lepton triplets at $y=0$ which is a $SU(3)_W$ symmetric
fixed point so that they  enjoy the $SU(3)_W$ symmetry. This also avoids possible
proton decay contact interactions.

One Higgs triplet $\mathbf{3}$ plus one Higgs anti-sextet
\sext, denoted as $\phi_6$,
with parities
\beqa
\phi_3(y)&=& {\rm P} \phi_3(-y)\;\;\;,  \phi_3(y')= {\rm P}' \phi_3(-y')\nonr\\
\phi_6(y)&=& {\rm P} \phi_6(-y){\rm P}^{-1}\;, \phi_6(y')=- {\rm P}' \phi_6(-y') {\rm P}'^{-1}.
\eeqa
is the minimal scalar set  to give viable charged fermion
masses (see \cite{SU3:triumf} ).
Another Higgs triplet \tripp with parities $(+-)$ is introduced
to transmit lepton number violation essential  for generating Majorana neutrino mass
through one-loop diagrams \cite{SU3:triumf}
by a triple Higgs interaction of  the type of
${\mathbf 3'}^T \bar{{\mathbf 6}}{\mathbf 3}$. This is a 5D realization of radiative
neutrino mass generation first proposed in \cite{Zee}. The resulting mass matrix
is necessarily of the Majorana type.

Now we have all the ingredients to write down explicitly the  5D Lagrangian density
\beqa
{\cal L}_5 &=& -\frac12 \TR[G_{MN}G^{MN}]
 +\TR[(D_M\phi_6)^\dag(D^M\phi_6)]\nonr\\
&+& (D_M\phi_3)^\dag(D^M\phi_3)
 +(D_M\phi'_3)^\dag(D^M\phi'_3)\nonr\\
&+& \delta(y) \left[\epsilon_{abc}\frac{ f_{ij}^3}{\sqrt{M^*}} \overline{(L^a_i)^\CC} L^b_j \phi_3^c
 + \epsilon_{abc}\frac{f_{ij}^{'3}}{\sqrt{M^*}}  \overline{(L^a_i)^\CC} L^b_j \phi_3^{'c}
\right.\nonr\\
&+& \left. \frac{f^6_{ij}}{\sqrt{M^*}}\overline{(L^a_i)^\CC} \phi_{6 \{ab\}} L^b_j
 + \bar{L}i\gamma^\mu D_\mu L\right]\nonr\\
&-& V_0(\phi_6,\phi_3,\phi'_3) -
 \frac{m}{\sqrt{M^*}}\phi_3^T \phi_6 \phi'_3 + H.c.\nonr\\
&+& {\cal L}_{\rm GF} + \mbox{quark sector}.
\label{5DL}
\eeqa
where $G_{MN}, M,N=\{0,1,2,3,y\}$ is the 5D field strength and $D_M$ is the 5D
covariant derivative.
The cutoff scale $M^*$ is
introduced to make the coupling constants  dimensionless.
The other notations are self explanatory.
The quark sector is not relevant now and will be left out.
The complicated scalar potential is gauge invariant and orbifold symmetric
and will not be specified since it is not needed here.
To perform loop calculations, we need to specify the
5D gauge fixing term, ${\cal L}_{\rm GF}$, which will be exhibited later.

The fields and their parities  of this model  are summarized below:
\beqa
8^\mu &=& \underbrace{(1,0)_{++}}_{B^\mu}
+\underbrace{(3,0)_{++}}_{A^\mu}
+\underbrace{(2,-3/2)_{+-}  + (2,+3/2)_{+-}}_{(U,V)^\mu}\nonr\\
8^y &=& \underbrace{(1,0)_{--}}_{B^y}
+\underbrace{(3,0)_{--}}_{A^y}
+\underbrace{(2,-3/2)_{-+}  + (2,+3/2)_{-+}}_{(U,V)^y}\nonr\\
\mathbf{3} &=& \underbrace{(2,-1/2)_{++}}_{H_{W1}}
+\underbrace{(1,1)_{+-}}_{H_{S}}\nonr\\
\mathbf{3}' &=& \underbrace{(2,-1/2)_{+-}}_{H'_{W1}}
+\underbrace{(1,1)_{++}}_{H'_{S}}\nonr\\
\mathbf{\bar{6}} &=& \underbrace{(3,+1)_{+-}}_{H_{T}}
+\underbrace{(2,-1/2)_{++}}_{H_{W2}}+\underbrace{(1,-2)_{+-}}_{H_{S2}}\nonr
\eeqa
where the SM  quantum numbers are $(SU(2)_L,U(1)_Y)$ and the subscripts
label the parities $P, P^{\prime}$.
Then it is straightforward to obtain the 4D effective interaction by
integrating over $y$ 
 and
the  4D effective gauge coupling can be identified as
$g_2= \tilde{g} / \sqrt{2\pi R M^*}$.
The orbifold construction is engineered such that there is no tree level LFV in the SM gauge interactions. Thus, the success of that model remains intact.
But the tree level LFV interaction emerge in the $U,V$ gauge
interaction which are heavy KK excitation and in the Yukawa interactions.

The LFV charged current is
\beqa
{\cal L}_{\rm CC} = g_2 \sum_{n=1}
\overline{e_{L i}}\gamma^\mu P_L ({\cal U}_{lep})_{ij}
e^\CC_{R j} U^{-2}_{n,\mu} +H.c.\nonr\\
+ g_2 \sum_{n=1}
\overline{\nu_{L i}}\gamma^\mu P_L ({\cal U}_{lep})_{ij}
e^\CC_{Rj} V^{-1}_{n,\mu} +H.c.
\eeqa
where the subscripts $L$ and $R$ are kept for book keeping.
The matrices $U_{L,R}$ are used to diagonalize the charged lepton mass
matrix and $ {\cal U}_{lep}=U_L^\dag U_R^* $ is an extra CKM-like
unitary mixing matrix for the lepton sector.

The LFV Yukawa interactions are  given by
\beqa
{\cal L}_{\rm Y}
&=& { 1 \over \sqrt{2\pi R M^*}}\sum_{n=0} \kappa_n  \left [
 f^3_{S ij} \overline{e^\CC_{L,i}}\nu_{L,j} H^+_{S,n}
+ f^3_{H ij}\left(\overline{e_{R,i}}e_{L,j} H^0_{W1,n}
+ \overline{e_{R,j}}\nu_{L,i} H^-_{W1,n}\right) -(i\Leftrightarrow j)
\right]\nonr\\
&+&
{1 \over \sqrt{2\pi R M^*}}\sum_{n=0}\kappa_n  \left [
f^{'3}_{S ij} \overline{e^\CC_{L,i}}\nu_{L,j} H^{'+}_{S,n}
+ f^{'3}_{H ij}\left( \overline{e_{R,i}}e_{L,j} H^{'0}_{W1,n}
+ \overline{e_{R,j}}\nu_{L,i} H^{'-}_{W1,n}\right) -(i\Leftrightarrow j)
\right]\nonr\\
&+&{1 \over \sqrt{2\pi R M^*}}\sum_{n=0} \kappa_n  \left [
 f^6_{T ij} \left(\overline{e^\CC_{L,i}}e_{L,j} H^{+2}_{T,n}
+( \overline{e^\CC_{L,i}}\nu_{L,j}
+ \overline{\nu^\CC_{L,i}}e_{L,j} ) H^+_{T,n}
+ \overline{\nu^\CC_{L,i}}\nu_{L,j} H^0_{T,n} \right)\right.\nonr\\
&+&\left. f^6_{H ij}\left( ( \overline{e_{R,i}}e_{L,j} + \overline{e_{L,i}}e_{R,j}) H^0_{W2,n}
+ (\overline{e_{R,i}}\nu_{L,j}+ \overline{\nu_{L,i}}e_{R,j}
)H^-_{W2,n}\right) +  f^6_{S ij} \overline{e_{R,i}} e^\CC_{R,j} H^{-2}_{S2,n}
\right] + H.c.
\eeqa
where $\kappa_n=(\sqrt{2})^{1-\delta_{n,0}}$ and
\beq
f^6_{T}= U_L^T f^6 U_L\,,\,f^6_{S}= U_R^T f^6 U_R\,,\, f^{(')3}_{S}= U_L^T f^{(')3} U_L\,,\,
 f^{(')3}_{H}= U_R^\dag f^{(')3} U_L\,,\, f^{6}_{H}= U_R^\dag f^{6}
 U_L\,.
\eeq
Note that in the new basis the symmetry of $f_T$ and $f_S$ are not changed.

\subsection{$L\ra l+\gamma$ transition}
\label{sec:SU3case}

We begin the discussion by studying a  special case that
$f_6 \gg f_3$, such that $U_R\sim U_L^*$ also $f^6_T, f^6_H$ and
$f^6_S$ are roughly diagonal.
This hierarchical Yukawa structure is also demanded to yield
the observed charged lepton mass hierarchy.
In other words, all the LFV sources are
in the Yukawa interaction of $\phi_3$ and $\phi_3'$.
Since $\phi'_3$ has nothing to do with the charged lepton masses,
we can further assume its LFV contribution is larger than $\phi_3$,
whose coupling is roughly $\sim (m/M_W)(f_3/f_6)$,
and $ f_S^{'3}\sim f_H^{'3}$.

In general this class of decays proceeds via the one-loop diagrams. The ones involving the
gauge boson $U^{\pm 2}$ and $V^\pm$ are suppressed by the GIM mechanism.
This leaves the singly charged and neutral  scalars
 as the only possible contributors since they both carry two units of
lepton charges in the usual scheme. We thus conclude that these decays
are dominated  by the scalar induced  M1 and E1 operators only.
Therefore, they provide  unique probes of the
exotic scalar sector. Later we will show that in contrast
$L\ra 3l$ probes the gauge interactions of the model.

\begin{figure}[htc]
  \centering
  \includegraphics[width=0.6\textwidth]{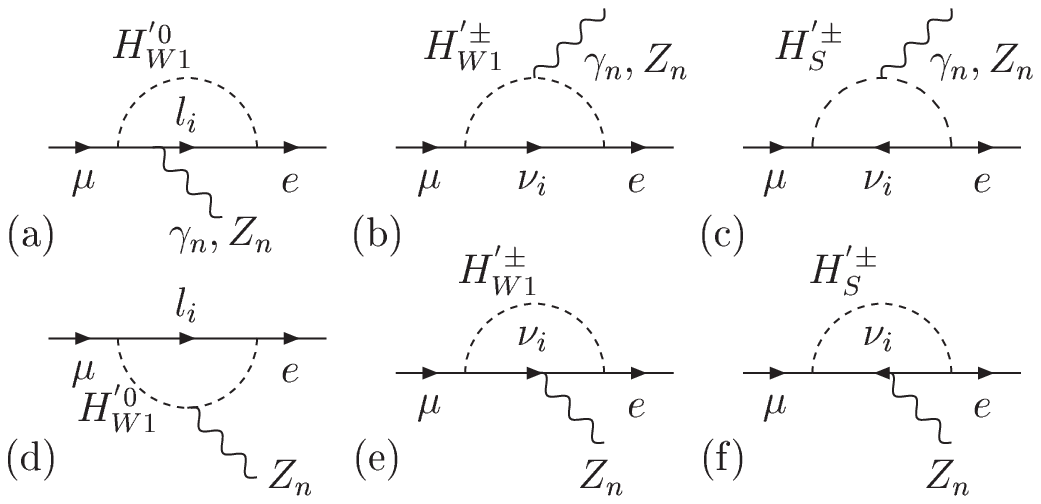}
  \caption{The leading  contributions for $\mu\ra e \gamma$ and $\mu-e$ conversion
  in the case described in SEC.\ref{sec:SU3case}.
  The labels $n\geq 0$ indicate the KK level.}
  \label{fig:muetransition}
\end{figure}

In this case, the leading contribution loop diagrams are shown in
Fig.\ref{fig:muetransition}.
Now, briefly discuss the gauge fixing in this model.
Because the orbifold parity for $\phi_3'$ is chosen to be $(+-)$,
it can not develop a VEV and doesn't participate the electroweak
breaking. The Goldstone bosons consist of  the
 $y$-components of gauge bosons and the proper linear combinations
of $\phi_3$ and $\phi_6$.
And the whole ${\mathbf 3}'$, $H_{W1}^{'0}, H_{W1}^{'\pm}$ and $H_S'$ are physical Higgs.
So now it is straightforward to carry out loop calculation.
 For further details, see Appendix.

The $E1, M1$ form factors are calculated to be:
\beqa
\label{eq:SU3M1}
f_{M1}^{L l}&=& {m_\mu^2 \over 384 \pi^2 M_{S0}^2}\sum_i
\left[ f_{Li} + {\epsilon  \over 24}(9f_{Ri}+7f_{Li}) \right]\\
\label{eq:SU3E1}
 f_{E1}^{L l}&=& {m_\mu^2 \over 384 \pi^2 M_{S0}^2}\sum_i
\left[ f_{Li} - {\epsilon  \over 24}(9f_{Ri}-7f_{Li}) \right]
\eeqa
where $f_{Li}=f_{Ri}^*\equiv f^*_{l i}f_{i L}$, $M_{S0}$ is the zero
mode mass of $H^{'\pm}_S$, and $\epsilon= (\pi M_{S0}  R)^2\sim{\cal O}(0.1)$.
On arriving at the above expression, the contributions of all KK scalar
excitation running in the loop have been summed.
And if we drop the $\epsilon$-terms,
the resulting branch ratio can  be expressed as:
\beqa
B(L\ra l+\gamma)&=& {96\pi^3 \alpha \over G_F^2 m_\mu^4}
\left(|f_{E1}^{Ll}|^2+|f_{M1}^{L l}|^2\right)
\sim {\alpha \over 768 \pi G_F^2 M^4_{S0}}\left|\sum_{i=e,\mu,\tau}
f_{Li}\right|^2\\
&=& 2.75\times10^{-6} \left({300 \mbox{GeV}\over M_{S0}}\right)^4
\left| (f^{'3}_{S,le})^*f^{'3}_{S,eL}+
       (f^{'3}_{S,l\mu})^*f^{'3}_{S,\mu L}+
       (f^{'3}_{S,l\tau})^*f^{'3}_{S,\tau L} \right|^2
\eeqa
Because the Yukawa couplings of triplet scalars are
anti-symmetric, the $L\ra l+\gamma$ processes have following forms:
\beqa
B(\mu\ra e+\gamma) = 2.75\times10^{-6} \left({300 \mbox{GeV}\over M_{S0}}\right)^4
\left| (f^{'3}_{S,e\tau})^*f^{'3}_{S,\mu\tau}\right|^2\\
B(\tau\ra e+\gamma) = 2.75\times10^{-6} \left({300 \mbox{GeV}\over M_{S0}}\right)^4
\left| (f^{'3}_{S,e\mu})^*f^{'3}_{S,\mu\tau}\right|^2\\
B(\tau\ra \mu+\gamma) = 2.75\times10^{-6} \left({300 \mbox{GeV}\over M_{S0}}\right)^4
\left| (f^{'3}_{S, e\mu})^*f^{'3}_{S,e\tau}\right|^2
\eeqa
We have taken $M_3'=300$GeV as the reference point.
If all of the Yukawa couplings  are real and none of them
vanishes,
their ratios can be further simplified to:
\beq
B(\mu\ra e+\gamma):B(\tau\ra e+\gamma):B(\tau\ra \mu+\gamma)
= {1\over |f^{'3}_{S,e\mu}|^2}:{1\over |f^{'3}_{S,e\tau}|^2}:{1\over
|f^{'3}_{S,\mu\tau}|^2}\,.
\eeq

At this point one can use the data  $B(\mu\ra e+\gamma) < 1.2 \times 10^{-11}$\cite{PDG} to obtain
the constrain $\left| (f^{'3}_{S,e\tau})^*f^{'3}_{S,\mu\tau}\right| <2.1 \times 10^{-3}$.
This is consistent with the expectation from the study of neutrino mass in this as given
in \cite{SU3:triumf}. There it was found that the Yukawa coupling $f^{'3}_{e\mu}$ has
to be  $\lesssim 10^{-2}$ and the $f$'s
 exhibit the  pattern $f^{'3}_{e\mu}>f^{'3}_{e\tau} >f^{'3}_{\mu\tau}$. Hence it reasonable
 $\mu\ra e+\gamma$ to occur at a rate less than two orders of magnitude below current level.
Indeed  in the $SU(3)_W$ model  we can link the  various $L\ra l\gamma$ transition branch ratios
to the light neutrino mass matrix elements.
Assuming that the light neutrino mass are mostly coming from the one-loop quantum correction involving
the zero modes of $\phi_3'$ and $\phi_6$, we have the prediction :
\beq
B(\mu\ra e+\gamma):B(\tau\ra e+\gamma):B(\tau\ra \mu+\gamma)
\sim  \left({m_\mu\over m_\tau}\right)^4 m_{13}m_{23}:
m_{12}m_{23}:m_{12}m_{13}
\eeq
where $m_{ij}$ is the $(ij)$ entry of the light neutrino mass matrix. Interestingly
 the model naturally accommodates an active neutrino mass matrix of the inverted hierarchy type as follows:
\beq
m\sim \left(\bray {ccc} \epsilon &1&1\\ 1&\epsilon  &\epsilon \\1&\epsilon &\epsilon^2
\eray \right)
\eeq
where $\epsilon \sim 0.1$.
From the above equations, we see  that $\mu\ra
e\gamma$ is suppressed compared to the $\tau\ra l \gamma$ decays. This is a striking feature of the
model.

\subsection{ $\mu-e$ conversion}
The $\mu-e$ conversion in nuclei will be dominated by the virtual photon exchange. Compared
to $\mu\ra e\gamma$ it has additional contributions from the anapole terms.
The corresponding photon $E0, M0$ form factors can be derived as:
\beqa
\label{eq:SU3E0}
\tilde{f}_{E0}(-k^2)&=& {m_\mu^2 \over 576 \pi^2 M_{S0}^2}
\sum_{i=e,\mu,\tau}\left[
 f_{Li} + {\epsilon  \over 24}( 3f_{Ri}+f_{Li} )
 -{6\epsilon\over \pi^2}(f_{Ri}+f_{Li})\sum_{n=1}^\infty {G(\delta_n, x_i)\over (2n-1)^2}
\right]\\
\label{eq:SU3M0}
\tilde{f}_{M0}(-k^2)&=& {m_\mu^2 \over 576 \pi^2 M_{S0}^2}
\sum_{i=e,\mu,\tau}\left[
 f_{Li} + {\epsilon  \over 24}( 3f_{Ri}- f_{Li} )
 -{6\epsilon\over \pi^2}(f_{Ri}-f_{Li})\sum_{n=1}^\infty { G(\delta_n, x_i)\over (2n-1)^2}
\right]
\eeqa
where $\delta_n=(-k^2)/M^2_{Hn}$, $x_i= m_i^2/(-k^2)$,  $i=e,\mu, \tau$, and
\beq
G(\delta, x)= -\ln \delta - \ln x
+\frac13-4x+(1-2x)\sqrt{1+4x}\ln{\sqrt{4x+1}-1\over\sqrt{4x+1}+1}.
\eeq
As expected, the principal contribution is from the Fig.\ref{fig:muetransition}(c)
with the  $H^{'\pm}_S$ zero mode running in the loop.
The logarithmic enhancements in $G(\delta,x)$, is due to the exchange of neutral scalars,
 $H_{W1}^{'0}$ ( see Fig.\ref{fig:muetransition}(a)). Although they  are suppressed by the KK masses
we find them to be  compatible to the charged singlet contribution.

In this  process $-k^2\sim m_\mu^2$ and $G$ has the following
limits
\beq
G_{e,n} \sim -\ln \frac{m^2_\mu}{M^2_{Hn}}+\frac13,\;
G_{\mu,n} \sim -\ln \frac{m^2_\mu}{M^2_{Hn}}-1.515,\;
G_{\tau,n} \sim -\ln \frac{m^2_\mu}{M^2_{Hn}}-6.978
\eeq
 The KK sum of these logarithmic enhancements
are finite:
\beq
\sum_{n=1}^\infty {1\over (2n-1)^2 }\ln \frac{m^2_\mu}{M^2_{Hn}}
= \frac{\pi^2}{8}\ln(m_\mu R)^2 -0.8362
\eeq
So the desired anapole form factors can be expressed as
\beqa
\tilde{f}_{E0}(m_\mu^2)&=& {m_\mu^2 \over 576 \pi^2 M_{S0}^2}
\sum_{i=e,\mu,\tau}\left[
 f_{Li} + {\epsilon  \over 24}( 3f_{Ri}+f_{Li} )
 +{3\epsilon\over 4}(f_{Ri}+f_{Li})[\ln(m_\mu R)^2+\eta_i ]
\right]\\
\tilde{f}_{M0}(m_\mu^2)&=& {m_\mu^2 \over 576 \pi^2 M_{S0}^2}
\sum_{i=e,\mu,\tau}\left[
 f_{Li} - {\epsilon  \over 24}( 3f_{Ri}- f_{Li} )
  +{3\epsilon\over 4}(f_{Ri}-f_{Li})[\ln(m_\mu R)^2+\eta_i ]\right]
\eeqa
$\{\eta_e, \eta_\mu,\eta_\tau\}=\{-1.011,0.837, 6.300\}$.
Again, since $f_S^{'3}$ is anti-symmetric, only $f_{L\tau}=f^*_{R\tau}$ can
contributes.
For simplicity  we assume there is no new CP violation in the scalar
sector; then $f_L=f_R$ and the $\mu-e$ conversion rate in $ ^{48}_{22}Ti$
can be expressed as:
\beq
\label{eq:MUEGSU31}
B^{\gamma}_{\rm conv}
\sim 0.01 B(\mu\ra e+\gamma)
\eeq
if taking $1/R=2$TeV and $M_S=300$GeV as a reference point.
It is also possible to have extra contributions from KK
photon and KK $Z$ excitation, Fig.\ref{fig:muetransition}(a-f).
One will need to take  care of the KK number conservation
in the scalar-scalar-gauge boson vertices and sum over all the
possible combinations. But generally speaking, their contributions
are further suppressed by $(m_\mu R)^2< 2\times 10^{-9}$
 compared to the photon zero mode and can be safely ignored.

The relation of Eq.(\ref{eq:MUEGSU31}) is based on the assumption
that $f_6\gg f_3$ and $\phi_3'$ is the dominate LFV source.
However, we should point out that if $f_3$ is not so small
the neutral scalar zero modes can make $\mu\ra e\gamma$ and
$B^{\gamma}_{\rm conv}$ compatible and deviates a lot from the pure
photonic dipole prediction, Eq.(\ref{eq:Weinberg}).

Again, this demonstrates that $L\ra l\gamma$ and $\mu-e$ conversion
are very important for us to understand the Yukawa structure in
the $SU(3)_W$ model.

The question now arises about the photonic dipole and anapole contribution to
$\mu\ra 3e$. The answer lies in  Eqs.(\ref{eq:SU3M1},\ref{eq:SU3E1},\ref{eq:SU3E0},
\ref{eq:SU3M0}). We estimated that
\beq
B(\mu\ra 3e)< 0.04 B(\mu\ra e\gamma).
\eeq
 This prediction is  not very sensitive
to what the Yukawa pattern is. Moreover, the decays $L\ra 3 l$ have overwhelming contribution
from other sources of new physics in the model to which we shall turn our attention to next.

\subsection{$L\ra 3 l$}

\begin{figure}[htc]
  \centering
  \includegraphics[width=0.8\textwidth]{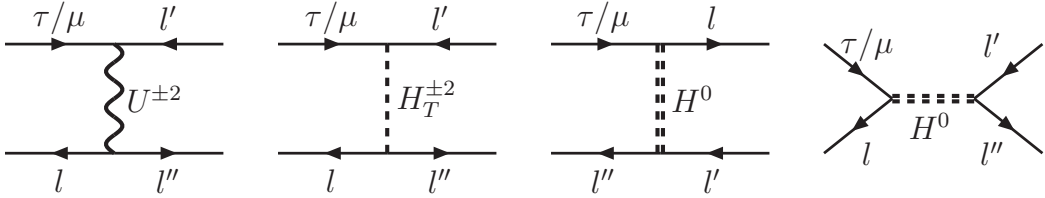}
  \caption{Feynman diagrams which lead to $\tau(\mu)\ra 3l$ processes. }
  \label{fig:su3_tau3l}
\end{figure}

A characteristic of the model is the existence of double charged gauge bosons with
LFV couplings. This will induce $\mu\ra 3e$ like processes for the $\tau$. In addition
there are also KK scalars $H_T$ and $H_0$ which has LFV Yukawa couplings which are largely unknown.
The Feynman diagrams for the $L\ra 3l$ decays are depicted in
Fig.\ref{fig:su3_tau3l}.
 Since the Yukawa coupling are totally unknown, we will postpone the discussion
 of the  contributions from scalars  and  look at
 the branch ratios, normalized to $B(\tau\ra e\nu_\tau\bar{\nu}_e)$,
 mediated by $U^{\pm 2}$ gauge boson alone first:
\beqa
B(\tau\ra 3 \mu)&=& {\cal F}\times
 \left(|{\cal U}_{\tau\mu }|^2+ |{\cal U}_{\mu\tau}|^2\right) |{\cal U}_{\mu\mu}|^2\,,\\
B(\tau\ra 3 e)&=& {\cal F}\times
 \left(|{\cal U}_{\tau e}|^2+ |{\cal U}_{e\tau}|^2\right) |{\cal U}_{ee}|^2\,,\\
B(\tau\ra  \bar{\mu} ee)&=&{\cal F}\times
 \left(|{\cal U}_{\tau \mu}|^2+ |{\cal U}_{\mu\tau}|^2\right)
|{\cal U}_{e e}|^2\,,\\
B(\tau\ra  \mu\mu \bar{e})&=&{\cal F}\times
 \left(|{\cal U}_{\tau e}|^2+ |{\cal U}_{e\tau}|^2\right)
|{\cal U}_{\mu \mu}|^2\,,\\
B(\tau\ra  \mu e\bar{e})&=& \frac{\cal F}8
 \left(|{\cal U}_{\tau e}|^2+ |{\cal U}_{e\tau}|^2\right)
\times\left(|{\cal U}_{e\mu}|^2+|{\cal U}_{\mu e}|^2\right)\,,\\
B(\tau\ra   e\mu \bar{\mu})&=& \frac{\cal F}8
 \left(|{\cal U}_{\tau \mu}|^2+ |{\cal
 U}_{\mu\tau}|^2\right)
\times\left(|{\cal U}_{e\mu}|^2+|{\cal U}_{\mu e}|^2\right)\,.
\eeqa
where ${\cal F}=(M_W\pi R)^4/16=1.56\times 10^{-5}( 2\mbox{TeV}/ 1/R )^4$.
From the analysis given in sec.II, we know all scalar operators give positive
contribution. So even though we know nothing about the Yukawa couplings,
we can still derive an interesting  lower bond from
the unitarity of ${\cal U}_{lep}$
\beq
\label{eq:tau3elowB}
B(\tau\ra 3 e) \geq {\cal F}\times
  |{\cal U}_{ee}|^2 \left(1-|{\cal U}_{ee }|^2 \right)\,
\eeq
for a given $1/R$.

If one wants to keep compactification scale $1/R$ low, say $\sim
1.5$ TeV, then we would require  $|{\cal U}_{ee}|$ to be either close
to zero or one. Furthermore, if we take the upper bound of $1/R < 5$ TeV derived from
unification seriously we obtain
\beq
B(\tau\ra 3 e) > 8.0\times 10^{-7}
  |{\cal U}_{ee}|^2 \left(1-|{\cal U}_{ee }|^2 \right)\, .
\eeq
On the other hand, if we assume that the bilepton  gauge boson exchange is the dominating  FCNC source,
another interesting upper bond can be derived:
\beq
B(\tau\ra 3 e) < \frac{\cal F}4= 3.9\times 10^{-6} \left({2 \mbox{TeV} \over 1/R}\right)^4
\eeq
 with $|{\cal U}_{ee }|=1/\sqrt{2}$ in Eq.(\ref{eq:tau3elowB}).
Actually, if all the LFV Yukawa couplings are  associated with $\phi_3'$ as discussed in
the previous two  subsections, the tree-level bilepton scalar contributions to $\tau\ra 3e$ vanish
due to the antisymmetry of the Yukawa couplings.    However, the
present experimental limit, $1-3 \times 10^{-7}$ \cite{Aubert:2003pc} will indicate that
the compactification radius is closer to the upper limit of $5 {\mathrm TeV}^{-1}$ for this
particular case.

\section{5D $SU(5)$ model}
The orbifold $SU(3)_W$ model discussed above has many interesting and novel features;
however, the fact that quarks and leptons have to be treated differently is an obstacle towards
complete unification. It's a natural attempt to further unify the quarks and leptons
in a larger GUT group. The simplest group for that is  $SU(5)$. Now all fermions are on equal footing
and can be clustered into 2 $SU(5)$ representations,i.e.
$\Psi_{\bar{5}}=\{d^c, L\}, \Psi_{10}=\{ Q, u^c, e^c\}$.

Similar to the $SU(3)_W$ model, the model is embedded in the background geometry of
 $S_1/Z_2\times Z_2'$
orbifold. The bulk $SU(5)$ gauge symmetry is broken to the SM by orbifold
parities , with parity matrices $diag\{+++++\}$ and $diag\{---++\}$
for $Z_2$ and $Z_2'$ transformations respectively. These are generalizations of the
$SU(3)_W$ case.

Since no right-handed neutrinos are added, neutrino masses can be generated
through quantum correction by using either
$\mathbf{10}$ or $\mathbf{15}$ bulk scalars
plus the $\mathbf{\overline{5'}(10/15)\bar{5}}$ interaction mandated by breaking to the SM
gauge group.
The orbifold parities of $\mathbf{10}$ or $\mathbf{15}$ bulk scalars
are determined to be $(++)$ by  considerations of proton decay.
They split into following components:
\beqa
  \mathbf{15}_s(++)
  =P_{15}\left(6,1,-\frac23\right)_{++}+T_{15}(1,3,1)_{++}+
C_{15}\left(3,2,\frac16\right)_{+-}\,,\nonr\\
  \mathbf{10}_a(++)
  =P_{10}\left(\bar{3},1,-\frac23\right)_{++}+S_{10}(1,1,1)_{++}+
C_{10}\left(3,2,\frac16\right)_{+-}\,.\nonr
\eeqa
A careful analysis shows that by using $\mathbf{15(10)}$ the resultant
neutrino mass matrix favor the normal(inverted) hierarchy \cite{SU5:triumf}.
It was also found that extra fine tuning  efforts were  needed to
obtain phenomenologically acceptable neutrino mass patterns
 by using  $\mathbf{10}$ alone;  so we will only discuss the case
which implements $\mathbf{15}$.

The $P$ components induce tree-level  $K^0-\bar{K}^0$ mixing.
To satisfy  the experimental constraints, it is required that
$M_{P}> 10^{5}$ GeV.
On the other hand, the two bulk Higgs in $ \mathbf{5, 5'}$ which are responsible
for the SM electroweak symmetry breaking share the same $(++)$ parities
as  $\mathbf{15}$. The brane Yukawa interaction term is easily constructed to be
\beq
{\cal L}_Y = \delta(y)\left[ {\tilde{f}^{15}_{ij} \over \sqrt{M^*/2}}
\overline{ \psi^{\{A\}\mathbf{c}}_{\bar{5}i} }\psi^{\{B\}}_{\bar{5}j} \phi^{\{AB\}}_{15} + H.c.
\right],
\eeq
 where $A,B$ are the $SU(5)$ symmetry indices. It can be seen that to contain
the  necessary LFV source to generate neutrino Majorana
masses.The neutrino mass matrix elements are proportional to
$({\cal M})^\nu_{ij}\propto \sum_k m_k f^{'5}_{ik} f^{15}_{jk}$
where $i,j,k$ are the generation indices and $m_k$ is the mass of $k$-charged
lepton running in the loop.

The extra Higgs doublet in the $\mathbf {5'}$ is good for gauge unification. By
adding additional decaplet bulk fermion pair with $(+-)$ parity
and mass around $10-120$ TeV, the unification is achieved at
$3\times 10^{16}-10^{15}$ GeV or equivalently $1/R \sim 10^{14}$ GeV.
The high scale unification or tiny radius of extra dimension makes
KK excitation decouple from most phenomenological studies  and  basically we  only
need to consider the zero modes.

Below unification scale or equivalently the low energy 4D effective theory is  a
two  Higgs doublets like model.  In general the two Yukawa
patterns are not aligned which that can  lead to severe tree
level charged neutral flavor changing (FCNC) interaction. A $Z_2$ symmetry is
usually assumed to forbid such tree level FCNC \cite{WG}.
In this model, there is no such freedom since the Yukawa patterns are determined by the
geometrical setup. The $\Psi_{10}$ of the first two
generations are assigned to be bulk fields and the other fermion
fields, $\Psi^3_{10}$ and $\Psi^{1,2,3}_{\bar{5}}$, are localized at
the $y=0$ brane. In  doing so, the salient
$SU(5)$ prediction of $m_b/m_\tau$ ratio is preserved  and give
small hierarchy  patterns  in the Yukawa couplings of  both $\mathbf{5, 5'}$ scalars, i.e.
\beqa
y_d \propto \left( \begin{array}{ccc} \delta & \delta& 1\\
\delta & \delta &1\\ \delta & \delta &1 \end{array}\right) ,
y_u \propto \left( \begin{array}{ccc} \delta^2 & \delta^2 &\delta\\
\delta^2 & \delta^2 &\delta\\ \delta & \delta &1 \end{array}\right).
\eeqa
Due to volume dilution factor we get  $\delta\sim 0.1$ which measures the
amount of overlap  between brane and bulk fields.
The specific Yukawa pattern  above successfully generates mass and mixing
hierarchy  of charged fermions:
\beqa
m_b:m_s:m_d= m_\tau: m_\mu:m_e \sim 1: \delta :\delta^2\\
m_t:m_c:m_u \sim 1:\delta^2 :\delta^4\,,\,
(V_{us}, V_{cb}, V_{ub})\sim (\delta, \delta, \delta^2)
\eeqa
The rotation from weak to mass eigenbasis simultaneously diagonalizes the  two Higgs doublets
Yukawa couplings. Thus, we do not have the  FCNC problem due to
mixing between two Higgs doublets.

\begin{figure}[htc]
  \centering
  \includegraphics[width=0.5\textwidth]{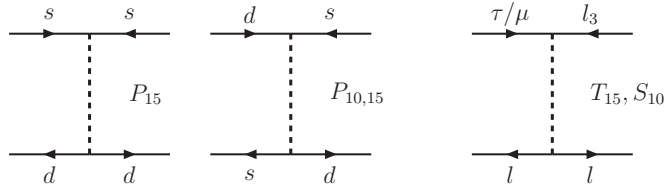}
  \caption{Feynman diagrams for tree level $K^0-\bar{K}^0$ mixing and $\mu\ra 3e$ process. }
  \label{fig:SU5_Ktau}
\end{figure}

Instead, now tree level  LFV processes can be  mediated by the triplet component $T_{15}$
in $\mathbf{15}$.  The only important ones are the   $\mu\ra 3e$ like processes, see
Fig.\ref{fig:SU5_Ktau}.

An explicit calculation give the branching ratio of $\mu\ra 3e$:
\beq
Br(\mu\ra 3e)
= {2|f^{15\dag}_{11} f^{15}_{12}|^2\over  g_2^4(\pi R M^*)^2}
\left( {M_W \over M_T}\right)^4.
\eeq
The mass difference $\tri M_K^P$ in $K^0-\overline{K^0}$ mixing arises from  $ P_{10,15}$ can be used to eliminate
the ambiguity of absolute strength of Yukawa couplings.
The ratio of Yukawa couplings can be replaced by the ratio of the
corresponding elements in $\cal {M}_{\nu}$. Since only the  $\mathbf{15}$ Higgs is used , we have
\beq
Br(\mu\ra 3e)\sim 3.02\times 10^{-16} \left({\tri m^P_K \over \tri m_K}\right)^2
\left(\frac{M_P}{M_T}\right)^4 \times
\left({2 m_{11}m_{12} \over m_{11}m_{22} + (2
\frac{m_e}{m_\mu}m_{12})^2}\right)^2.
\eeq
It is  straightforward to extend the analysis  to  $\tau\ra 3l$ transitions.
Assuming that the hierarchy of the elements of neutrino mass matrix is
smaller than factor 100, this model predicts
\beqa
Br(\mu\ra3e):Br(\tau\ra 3e):Br(\tau\ra3\mu):
Br(\tau\ra\mu e e):Br(\tau\ra e\mu\mu)\nonr\\
\sim \frac{m_{12}^2}{m_{22}^2}:
\left(\frac{m_\mu}{m_\tau}\right)^4\frac{m_{13}^2}{m_{22}^2}:
\left(\frac{m_e}{m_\tau}\right)^4\frac{m_{23}^2}{m_{11}^2}:
\left(\frac{m_\mu}{m_\tau}\right)^4\frac{m_{23}^2}{m_{22}^2}:
\left(\frac{m_e}{m_\mu}\right)^4\frac{m_{12}^2}{m_{11}m_{22}}.
\eeqa
The photonic form factors due to $T^{\pm2}, T^\pm$ scalar one loop diagrams
can be obtained:
\beqa
f_{M1}^{\mu e}=f_{E1}^{\mu e} &=&
-{m_\mu^2 \over 16 \pi^2 M_{T}^2} {5 f^{15}_{i\mu}(f^{15}_{ie})^*\over 24}\\
\tilde{f}^{\mu e}_{M0}= \tilde{f}^{\mu e}_{E0}&=&{m_\mu^2 \over 16 \pi^2 M_{T}^2}
{f^{15}_{i\mu}(f^{15}_{ie})^*  \over 6}\left[ G(\delta, x_i)-\frac12 \right]
\eeqa
The chiral structure in the above result is easily understood because that
 only the lepton doublets interact with triplet scalar.

Due to the factor $(m_\mu/M_T)^4$, the
$\mu\ra e \gamma$ process is strongly suppressed. Taking $M_T=10^5$GeV,
the  $\mu Ti \ra e Ti$ conversion rate is estimated to be
$\sim 1.2\times 10^{-14} |f^{15}_{i\mu}(f^{15}_{ie})^*|^2$
where we have kept only  logarithmic terms which is sufficient for an order of
magnitude estimate. The experimental bound is $3.6 \times 10^{-11}$ \cite{PDG}
which is not stringent. On the other hand the recently proposed experiment aimed at
a detecting a signal at the $10^{-16}- 10^{-17}$ will be very encouraging. Even
a negative result will provide stringent constrain on the otherwise unknown Yukawa
couplings.

\section{Split Fermion Model}

An interesting scenario was introduced by \cite{AS} to solve
the charged fermion masses hierarchy problem.
The basic idea is to postulate that fermions are bulk fields and they
interact with a non-dynamic background scalar potential.  In the 5D version the bulk
fermions are vectorlike, but only one of the chiral  zero modes
will be localized at the zero of the background potential modulated by the 5D mass terms.
The chirality of the zero mode is determined by the sign of slope of the
background potential at the zero point.
The fermion zero modes are given  a Gaussian profile in the fifth
dimension and each has its own unique position in the extra dimension.
The widths are controlled by the potential slope at the localized position.
For simplicity, we will assume a universal width
for all the SM fermions.
The 5D fermion excitation are vectorlike and will be located at
 the same position of  their zero modes. Roughly speaking,
the  energy gap is $\sim 1/(\mbox{Width})\gg 1/R \gg M_W$ and they
decouple from the phenomenology we are interested in.
Since the SM fermions are scattered over the fifth dimension, to
preserve the gauge interaction universality the SM gauge fields are
forced to be bulk fields too. To illustrate the basic physics involved
it suffices to build a model on an $S_1/Z_2$ orbifold so that one can
remove the unwanted $y$-components of gauge bosons zero modes which are
identified with the SM gauge bosons. However,
to break electroweak symmetry, a dynamic bulk Higgs is necessary.

To be more concrete we take the extra dimension to be the interval
$y\in[-\pi R, \pi R]$ and the fermions are fixed in different positions $z_i$ in this
interval. The usual Kaluza-Klein ansatz is invoked that any  bulk  field factorizes into
a 4D field
times a 5D wave function. Thus, for the fermion field we have $ \psi_i(x,y) = g(z_i,y) \psi(x) $
where $g(z_i,y)$ is taken to be  Gaussian distribution in the fifth
dimension:
\[
g(z_i, y)= {1 \over(\pi \sigma_G^2)^{1/4}} \exp\left[ -{(y-z_i)^2\over
2\sigma_G^2}\right]\]
where $\sigma_G$ is the universal width of Gaussian distribution.
If $\sigma_G\ll R$, $g$ acts like the Dirac delta function.
effective 4D interactions are obtained by integrating out the $y$ direction. Then any pair fermions get an
exponential suppression
\[
g(z_1,y) g(z_2,y) = \exp\left[ -{(z_1-z_2)^2\over 4\sigma_G^2}\right]
g\left(\frac{z_1+z_2}{2},y\right)
\]
as a result of integrating over Gaussian functions.
Therefore, the  linear displacement between left-handed and right-handed fermions
in the fifth dimension translates into  exponential Yukawa hierarchy in 4D theory.
One set of solution for the quarks positions have been found
numerically that can  accommodate the mass hierarchy and the CKM mixing\cite{Mirabelli:1999ks}.
To accommodate the CP violation, the overall Yukawa coupling
strength of up and down type quarks must be different\cite{Branco:2000rb,Chang:2002ww}.
Although a fundamental theory of where to place the fermions are located  is
lacking, we have at least one realistic solution  for where the quarks are located in the extra dimension.
For the lepton sector, we don't have enough constraints to pin down
the solution.
As pointed out by\cite{Chang:2002ne,Chang:2002ww}, the FCNC interaction is induced geometrically
where  phenomenological constraint in quark sector have also been discussed.
In fact, flavor violation  is   generic in any  multi-position models. This comes from
 the fact that weak to mass eigenbasis rotations can only make the SM interactions
(zero modes) flavor diagonal. the KK modes cannot be simultaneously diagonalized.

In general, the LFV can  be discussed independently of the neutrino sector.
The 4D effective LFV lagrangian can be expressed as
\beqa
{\cal L}^{LFV} &=& \sum_n \left({\kappa_n g_2\over \cos\theta}\right)
\bar{l}_i \gamma^\mu\left[ g_L U^L_{n(ij)}\hat{L} +g_R U^R_{n(ij)}\hat{R} \right] l_j
Z_\mu^n\nonr\\
&-& \sum_n (\kappa_n e)
\bar{l}_i \gamma^\mu\left[ U^L_{n(ij)}\hat{L} +U^R_{n(ij)}\hat{R}\right] l_j
A_\mu^n\nonr\\
&+&\sum_n \left({\kappa_n g_2\over \sqrt{2} }\right)
\bar{l}_i \gamma^\mu U^L_{n(ij)}\hat{L}\nu_j W_\mu^{n-} +h.c.
\eeqa
where $A^n, Z^n$ and $W^n$ are the n-th KK excitation of the
photon, $Z$ and $W$ bosons and $g_{L/R}= T_3(l)-Q_l \sin^2\theta_W$.
The matrices $U^{L/R}_n$ are a combination of the unitary
transformations $V_{L/R}$ that take the lepton weak eigenstates to
their mass eigenstates and the cosine weighting of the n-th KK
modes. Explicitly, they are
\beq
U^{L/R}_n =V_{L/R}^\dag
 \mbox{diag}\left(\cos\frac{n y_1^{L/R}}{R},
 \cos\frac{n y_2^{L/R}}{R},\cos\frac{n y_3^{L/R}}{R} \right) V_{L/R}
\eeq
for $n=0$ all the cosine factors become one and
 the interactions reduce to SM as expected.
It is clear that the $U^{L/R}$ is no more diagonal nor unitary in
general.

\begin{figure}[tc]
  \centering
  \includegraphics[width=0.5\textwidth]{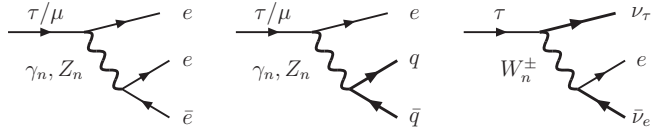}
  \caption{Flavor changing due to KK gauge bosons. }
  \label{fig:LFV_SF}
\end{figure}

In this  model, the $\tau\ra 3l$ decay and $\mu-e$ conversion happen at
tree-level, see Fig.\ref{fig:LFV_SF}.
The corresponding effective LFV couplings are:
\beqa
g_3&=& (2M_W R)^2 (\sin^2\theta_W+ g_R^2/\cos^2\theta_W)
\sum_{n=1} {U^{R*}_{n,ei}U^R_{n,i\mu}\over n^2}\\
g_4&=& (2M_W R)^2 (\sin^2\theta_W+ g_L^2/\cos^2\theta_W)
\sum_{n=1} {U^{L*}_{n,ei}U^L_{n,i\mu}\over n^2}\\
g_5&=& (2M_W R)^2 (\sin^2\theta_W+ g_R g_L/\cos^2\theta_W)
\sum_{n=1} {U^{R*}_{n,ei}U^L_{n,i\mu}\over n^2}\\
g_6&=& (2M_W R)^2 (\sin^2\theta_W+ g_L g_R/\cos^2\theta_W)
\sum_{n=1} {U^{L*}_{n,ei}U^R_{n,i\mu}\over n^2}
\eeqa
and  the $v, a, v_q$ and $a_q$ can be obtained in a similar way.
Also the lepton universality is broken due to flavor dependent couplings
in the KK gauge interaction. We refer the reader to \cite{Chang:2002ne} for a detailed analysis.

The leading $\mu\ra e \gamma$ contribution comes  from  the one loop corrections.
We need to fix the gauge before proceeding.
The necessary details of 5D gauge fixing are collected in the appendix.
After properly identifying the Goldstone boson, the usual 4D
$R_\xi$ gauge technique can be straightforwardly applied here.
Note that the Yukawa couplings of the physical KK scalars are suppressed by
the factor of $\sim (m_l R)/n$. Although there is residual GIM cancellation
in the KK gauge boson interaction, we expect the leading
LFV are from KK gauge interaction.

The LFV photonic form factors due to KK gauge boson and their
Goldstone boson can be calculated. The KK $W$ bosons' contribution to  the
 photonic form factors are given by:
\beqa
f^W_{M1}&=&f^W_{E1}={7\over 24}{g_2^2\over 16\pi^2}\sum_{n=1}
{(m_\mu R)^2\over n^2}  U^{L*}_{n,ie}U^L_{n,i\mu},\\
\widetilde{f^W_{M0}}&=&\widetilde{f^W_{E0}}={23\over 72}{g_2^2\over
16\pi^2}\sum_{n=1} {(m_\mu R)^2\over n^2}
U^{L*}_{n,ie}U^L_{n,i\mu}
\eeqa
and for the KK $Z$ bosons they are
\beqa
f^Z_{M1/E1}&=&{(m_\mu R)^2\over 8\pi^2} {g_2^2 \over\cos^2\theta
}\sum_{n=1}\sum_i {1\over n^2}\left\{ -\frac13 \left[
g_L^2 U^{L*}_{n,ie}U^L_{n,i\mu}\pm g_R^2
U^{R*}_{n,ie}U^R_{n,i\mu}\right]\right.\nonr\\
&& \left.+\frac{m_i}{m_\mu}g_L g_R\left[ U^{L*}_{n,ie}U^R_{n,i\mu}\pm
U^{R*}_{n,ie}U^L_{n,i\mu}\right] \right\}
 ,\\
\widetilde{f^W}_{E0/M0}&=& -{(m_\mu R)^2\over 24\pi^2} {g_2^2 \over\cos^2\theta
}\sum_{n=1}\sum_i {1\over n^2}\left\{ \left( G(\delta_n, x_i)+ \frac12\right)
\left[ g_L^2 U^{L*}_{n,ie}U^L_{n,i\mu}\pm g_R^2 U^{R*}_{n,ie}U^R_{n,i\mu}\right]
 \right\}
\eeqa
 Similar contributions from KK photons can be easily read from the above
 by replacing $(g_2/\cos\theta)\ra e$, $g_L\ra 1$, and  $g_R\ra 1$.

These photonic form factors give extra contribution to $\mu\ra
3e$ and  $\mu-e$ conversion processes but can't compete with those
tree-level KK gauge boson exchanging diagrams.
However they are the sole sources of new physics for the $L\ra l+\gamma$ process.

In addition to the usual ignorance with regard to Yukawa coupling there are more
unknowns in the lepton locations and the Gaussian widths.  Ad hoc simplifying
assumptions have to be made. Hence,
this kind of model suffers from a lack of predictive power in LFV studies.
More data such as the scale of neutrino mass and more complete knowledge
of the neutrino mixing matrix will help greatly.
However, we can extract  some generic features  for this kind of models as follow:
\begin{enumerate}
\item
 $\mu/\tau \ra 3l$ and $\mu Ti\ra e Ti$(or $\tau\ra l +{\mathrm {hadrons}}$)
will happen  at tree-level
from the  exchange of KK scalars, photons and $Z$ bosons.
\item  $L\ra l \gamma$ proceeds  at the one-loop level and hence is expected to be suppressed
compared to the previous modes
\item Violation of lepton universality will occur. The best signal will be to look for the
violation in $W\ra l_i\nu_i$ decays \cite{Chang:2002ne}.
\end{enumerate}
Unfortunately a more quantitative statement about the level of the effect eludes us for now.

\section{Conclusion}
We have studied and reviewed LFV processes in 5D gauge models that are related to neutrino
mass generation or address the flavor problem. Specifically we focus on two complete
models which generate neutrino masses radiatively. This allows us to see in
detail how the two issues can be related. The models are based on $SU(3)_W$ and $SU(5)$
5D unification. They give rise to different neutrino mass patterns
\cite{SU3:triumf,SU5:triumf};
 thus, it is not surprising that they give different prediction for LFV. The $SU(3)_W$ model
has a unification scale at $\sim {\mathrm TeV}$ and makes essential use of bileptonic
scalars. It also contains characteristic doubly charged gauge bosons. The $SU(5)$ model
is a 5D orbifold version of the usual GUT. The unification scale is much higher at $10^{15}$ GeV.
The important ingredient for LFV  and neutrino masses is the $\mathbf 15$ Higgs representation.
The triplet Higgs of this model plays the crucial role here.

We found that for the $SU(3)_W$ model the rare $\tau$ decays are much more enhanced
compare to their counterpart $\mu$ decays. Among the $\tau\ra l+\gamma$ decays the largest
mode is the $\mu +\gamma$. Even for this mode we expect it to be $<10^{-14}$ which
is much lower then current experimental reach.

The decay modes $\tau \ra 3l$ have  a better chance of being observe. This stems from the
fact that they are tree level processes induced by the bileptonic gauge bosons or scalars.
Since they are KK modes they have high masses controlled by the extra dimension compactification radius
which is $\leq 5 {\mathrm TeV}$ from consistency and unification considerations. An order
of magnitude
improvement on the current limit will be valuable information on the unknown Yukawa
couplings.

For the orbifold 5D $SU(5)$ model the muon to electron conversion in nuclei can be within the
experimental capability of the proposed experiment at Brookhaven National Laboratory\cite{Popp:2001hu}. As
in the previous model $\mu\ra e+\gamma$ will not be observable. This is very different
conventional 4D unificational models.

The split fermion model also have the characteristic of $L\ra 3l$ and $\mu\ra e$ conversion
dominating over $L\ra l\gamma$. We cannot be more quantitative due to proliferation of
unknown parameters. This model have lepton universality violation which is not present in
the previous two models. This can serve as a differentiating tool.

It is clear that in order to unravel the physics behind the flavor problem all modes
of LFV must be searched for. The usual 4D supersymmetric model will favor $L\ra l\gamma$
where as the 5D models prefer $L\ra 3l$ and/or $\mu \ra e$ conversion. To this we add
lepton universality test as a probe.

\acknowledgments{
WFC thanks Y Okada for helpful discussion and is grateful to Institute of
Theoretical Physics, Chinese Academy of Sciences,
for their kind hospitality where part of the work has been completed. This
work is also  supported in part by the Natural Science and Engineering Council of
Canada.}

\appendix
\section{$SU(2)\times U(1)$ in 5D $S_1/Z_2$ model with  a brane at $y=0$}
Now we present the gauge fixing scheme used for the 5D electroweak
interaction with one bulk Higgs doublet. For simplicity the background
geometry is $S_1/Z_2$.
The fifth gamma matrix was chosen to be $\gamma^y=i\gamma^5$. The 5D Lagrangian is
\beqa
{\cal L}_5 &=& -\frac14 F^{MN}F_{MN}-\frac14 G^{(a),MN}G^{(a)}_{MN}
+\left(D_M \Phi\right)^\dag \left(D^M \Phi\right)
+\cdots \label {eqn:L5}
\eeqa
where
\beqa
F_{MN}&=& \partial_M B_N - \partial_N B_M\,,\nonr\\
G^{(a)}_{MN}&=& \partial_M A^{(a)}_N - \partial_N A^{(a)}_M
+{\tilde{g}_2\over \sqrt{\MS}} \epsilon^{abc}A_M^{(b)} A_N^{(c)}\,\nonr\\
D_M &=&\partial_M -i {\tilde{g}_2\over \sqrt{\MS}} \frac{\tau^{(a)}}{2}A_M^{(a)}.
- i {\tilde{g}_1 Y\over \sqrt{\MS}} B_M\nonr
\eeqa
 $B$ and $A$  stand for the $U(1)$ hyper charge and $SU(2)$
gauge fields respectively. In this convention, $Q=T_3+Y$.
We adopt the usual conventions: $W^M_{\pm}=\frac{1}{\sqrt{2}}(A_1^M\mp iA_2^M)$,
 $P^M( \mbox{hoton})=(c_W B^M+ s_W A_3^M)$ and
$Z^M=( c_W A_3^M- s_W B^M)$, or $B^M=c_W P^M-s_W Z^M, A_3^M=c_W Z^M+s_W P^M$),
 where $c_W=\tilde{g}_2/\sqrt{\tilde{g}_1^2+\tilde{g}_2^2}$
and  $s_W=\tilde{g}_1/\sqrt{\tilde{g}_1^2+\tilde{g}_2^2}$.
 $\tilde{g}_5=\sqrt{\tilde{g}_1^2+\tilde{g}_2^2}$  are
introduced to simplify the notation.
The symmetry breaking pattern is same as in the usual 4D SM.
The bulk Higgs doublet acquires a nonzero VEV after SSB,
\beq
\Phi=\left(\begin{array}{c}   h^+ \\
{ \VEVB+ h^0\over \sqrt{2}} \end{array}\right)\,\,,\,
h^0=\phi^0+i\chi^0\, .
\eeq
 The generalized linear $R_\xi$ gauge fixing is introduced \cite{GaugeFixing}(other schemes can be found in
 \cite{GFCoV}),
\beqa
{\cal L}_{GF}=
&-&\frac{1}{2\xi} \left(\partial_\mu P^\mu +\xi \partial_y P^y  \right)^2
\nonr\\
&-& \frac{1}{\xi}\left|\partial_\mu W^{+,\mu}
+ \xi\left[\partial_y W^{+,y} - i M_W h^+\right]\right|^2
\nonr\\
&-& \frac{1}{2\xi}\left(\partial_\mu Z^\mu
+\xi\left[\partial_y Z^{y} - M_Z \chi^0\right] \right)^2.
\eeqa
to remove the mixing between gauge bosons and Higgs.
Therefore,  the Goldstone bosons and the
physical scalars can be easily identified:
\beqa
G^\gamma_n &=& P^y_n\,,\, M_\gamma^{(n)}= n/R\\
G^0_n &=& \left[c^Z_n Z_n^y - s^Z_n \chi^0_n\right]\,,\,
S^0_n= \left[s^Z_n Z_n^y + c^Z_n \chi^0_n\right]\nonr\\
 &&M_Z^{(n)}= \sqrt{n^2/R^2+M_Z^2}\,,\, s^Z_n= M_Z/M_Z^{(n)}\\
G^\pm_n&=& \left[c^W_n W_n^{y\pm} \mp i s^W_n h^\pm_n\right]\,,\,
H^\pm_n= \left[s^W_n W_n^{y\pm} \pm i c^W_n h^\pm_n\right]\nonr\\
 && M_W^{(n)}= \sqrt{n^2/R^2+M_W^2}\,,\, s^Z_n= M_W/M_W^{(n)}\\
H^0_n &=& \phi^0_n\,,\, M_H^{(n)}=\sqrt{n^2/R^2+M_\phi^2}
\eeqa
where $G^0$ and $G^\pm$ are the KK Goldstone bosons,
$S^0$ is the physical KK pseudo scalar, and $H^0, H^\pm$ are the
physical KK scalars.
The usual $R_\xi$ gauge can be extended to the 5D $S_1/Z_2$ model
with little modification, like $M_W\Rightarrow M_W^{(n)}$ and so on.

The Goldstone bosons are mainly constituted by the fifth gauge
components with a small fraction of KK Higgs bosons mixed interaction. On the other hand
 the Goldstone bosons
couple to brane fermions through  their Higgs components. In contrast the
physical scalars are mainly composed of  KK Higgs plus small amount
of the fifth components of gauge fields.

This scheme can also be applied to the models built on the $S_1/(Z_2\times Z'_2)$
orbifold with little modification.

\section{Gauge fixing for the orbifold models with more than one scalars.}
The method can be easily extended to the cases with multi scalars.
Taking a 5D two Higgs doublets Model(2HDM) as an example, with VEVs
$\langle\phi_1\rangle=v_1$, $\langle\phi_2\rangle=v_2$ and $\tan\beta=v_2/v_1$,
the physical charged scalars and pseudoscalars are
\beq
H^\pm = \sin\beta \phi_1^\pm - \cos\beta \phi_2^\pm\,,\,
A^0 = \sin\beta Im\phi_1^0 - \cos\beta Im \phi_2^0
\eeq
just like the usual 4D 2HDM. The only difference is that the
orthogonal linear combinations $ a^0 =  \cos\beta Im\phi_1^0 + \sin\beta Im\phi_2^0$
and $g^\pm=\cos\beta \phi_1^\pm + \sin\beta \phi_2^\pm$ will mix with the
fifth components of gauge fields to form the real Goldstone bosons:
\beqa
G^0_n &=& \cos\theta^0_n V^0_n- \sin\theta^0_n a^0_n \,,\,
\sin\theta^0_n= M_0 / \sqrt{M_0^2 +n^2/R^2}\, ,\\
G^\pm_n &=& \cos\theta^\pm_n V^\pm_n \mp i \sin\theta^\pm_n
g^\pm_n\,,\,
\sin\theta^\pm_n= M_V / \sqrt{M_V^2 +n^2/R^2}\, .
\eeqa


\begin{thebibliography}{99}

\bibitem{SuperK}
Y.~Fukuda {\it et al.}  [Super-Kamiokande Collaboration],
\prl{81}{1998}{1562}
[hep-ex/9807003].

\bibitem{SNO}
Q.~R.~Ahmad {\it et al.}  [SNO Collaboration],
\prl{89}{2002}{011301}
[nucl-ex/0204008].

\bibitem{KamL}
K.~Eguchi {\it et al.}  [KamLAND Collaboration],
\prl{90}{2003}{021802}
[hep-ex/0212021].


\bibitem {WMAP}
C.~L.~Bennett {\it et al.},
[astro-ph/0302207].

\bibitem {susylfv}
A.~Masiero, S.K.~Vempati, and O.~Vives, Nucl. \ Phys. \ B {\bf 649}, 189 (2003)\\
T.~Blazek and S.F.~King, {\it ibid} {\bf 662}, 359 (2004)

\bibitem{CN04}
W.F.~Chang and J.N.~Ng [archiv:hep-ph/0411201]

\bibitem{SU3:triumf}
C.~H.~Chang, W.~F.~Chang and J.~N.~Ng,
Phys.\ Lett.\ B {\bf 558}, 92 (2003)
[arXiv:hep-ph/0301271].


\bibitem{SU5:triumf}
W.~F.~Chang and J.~N.~Ng,
JHEP {\bf 0310}, 036 (2003)
[arXiv:hep-ph/0308187].
\bibitem{JN}
J.N.~Ng, J. Korean \ Phy. \ Soc. {\bf 45} S341 (2004)

\bibitem{Okada}
Y.~Kuno and Y.~Okada,
Rev.\ Mod.\ Phys.\  {\bf 73}, 151 (2001)
[arXiv:hep-ph/9909265] and references therein.

\bibitem{Petcov}
S.~T.~Petcov,
Phys.\ Lett.\ B {\bf 68}, 365 (1977).

\bibitem{Huitu}
K.~Huitu, J.~Maalampi, M.~Raidal and A.~Santamaria,
Phys.\ Lett.\ B {\bf 430}, 355 (1998)
[arXiv:hep-ph/9712249].\\
T.~S.~Kosmas, S.~Kovalenko and I.~Schmidt,
Phys.\ Lett.\ B {\bf 511}, 203 (2001)
[arXiv:hep-ph/0102101].

\bibitem{weinberg}
G.~Feinberg and S.~Weinberg,
Phys.\ Rev.\ Lett.\  {\bf 3}, 527 (1959).


\bibitem{Kitano:2002mt}
R.~Kitano, M.~Koike and Y.~Okada,
Phys.\ Rev.\ D {\bf 66}, 096002 (2002)
[arXiv:hep-ph/0203110].

\bibitem{Chiang:1993xz}
H.~C.~Chiang, E.~Oset, T.~S.~Kosmas, A.~Faessler and J.~D.~Vergados,
Nucl.\ Phys.\ A {\bf 559}, 526 (1993).


\bibitem{Wein75}
S.~Weinberg, Phys. \ Rev. \ D {\bf 2}, 1962 (1975)
\bibitem{5DSU3}
L.J.~Hall and Y.~Normura, Phys. \ Lett. \  B {\bf 532} 111 (2002)\\
T.~Li and W.~Liao, {\it ibid } {\bf 545} 147 (2002)\\
S.~ Dimopoulos,D.E.~Kaplan, and N.~Weiner, {\it ibid} {\bf 534} 124 (2002)\\
I.~Antoniadis and K.~Benakli, {\it ibid} {\bf 326} 69 (1994)

\bibitem{Zee}
A.~Zee, Phys. \ Lett. \ B {\bf 161} 141 (1985)

\bibitem{PDG}
Particle Data Group, Phys. \ Lett. B {\bf 592} 410 (2004)

\bibitem{Aubert:2003pc}
B.~Aubert {\it et al.}  [BABAR Collaboration],
Phys. \ Rev. \ Lett. {\bf 92} 121801 (2004)



\bibitem{WG}
S.~Glashow and S.~Weinberg, phys. \ Rev. \ D {\bf 15} 1958 (1977).

\bibitem{AS}
N.~Arkani-Hamed and M.~Schmaltz,
Phys.\ Rev.\ D {\bf 61}, 033005 (2000)
[arXiv:hep-ph/9903417];\\
N.~Arkani-Hamed, Y.~Grossman and M.~Schmaltz,
Phys.\ Rev.\ D {\bf 61}, 115004 (2000)
[arXiv:hep-ph/9909411].

\bibitem{Mirabelli:1999ks}
E.~A.~Mirabelli and M.~Schmaltz,
Phys.\ Rev.\ D {\bf 61}, 113011 (2000)
[arXiv:hep-ph/9912265].

\bibitem{Branco:2000rb}
G.~C.~Branco, A.~de Gouvea and M.~N.~Rebelo,
Phys.\ Lett.\ B {\bf 506}, 115 (2001)
[arXiv:hep-ph/0012289].

\bibitem{Chang:2002ww}
W.~F.~Chang and J.~N.~Ng,
JHEP {\bf 0212}, 077 (2002)
[arXiv:hep-ph/0210414].


\bibitem{Chang:2002ne}
W.~F.~Chang, I.~L.~Ho and J.~N.~Ng,
Phys.\ Rev.\ D {\bf 66}, 076004 (2002)
[arXiv:hep-ph/0203212].

\bibitem{Popp:2001hu}
J.~L.~Popp  [MECO Collaboration],
Nucl.\ Instrum.\ Meth.\ A {\bf 472}, 354 (2000)
[arXiv:hep-ex/0101017].

\bibitem{GaugeFixing}
D.~M.~Ghilencea, S.~Groot Nibbelink and H.~P.~Nilles,
Nucl.\ Phys.\ B {\bf 619}, 385 (2001)
[arXiv:hep-th/0108184];\\
A.~Muck, A.~Pilaftsis and R.~Ruckl,
Phys.\ Rev.\ D {\bf 65}, 085037 (2002)
[arXiv:hep-ph/0110391].

\bibitem{GFCoV}
For example,
K.~R.~Dienes, E.~Dudas and T.~Gherghetta,
Nucl.\ Phys.\ B {\bf 537}, 47 (1999)
[arXiv:hep-ph/9806292];\\
J.~Papavassiliou and A.~Santamaria,
Phys.\ Rev.\ D {\bf 63}, 125014 (2001)
[arXiv:hep-ph/0102019];


\end{thebibliography}
\end{document}